\documentclass{article}

\usepackage{graphicx}  
\usepackage{subfigure} 
\usepackage{dcolumn}   
\usepackage{bm}        
\usepackage{amsmath}   
\usepackage{soul}
\usepackage[margin=1in]{geometry}
\usepackage{hyperref}

\begin{document}

\title{Implications of a 20-Hz Booster cycle-rate for Slip-stacking}
\author{Jeffrey Eldred \& Robert Zwaska}
\date{\today}

\maketitle

\begin{abstract}
We examine the potential impacts to slip-stacking from a change of the Booster cycle-rate from 15- to 20-Hz. We find that changing the Booster cycle-rate to 20-Hz would greatly increase the slip-stacking bucket area, while potentially requiring greater usage of the Recycler momentum aperture and additional power dissipation in the RF cavities.   In particular, the losses from RF interference can be reduced by a factor of 4-10 (depending on Booster beam longitudinal parameters). We discuss the aspect ratio and beam emittance requirements for efficient slip-stacking in both cycle-rate cases.  Using a different injection scheme can eliminate the need for greater momentum aperture in the Recycler.
\end{abstract}

\section*{Introduction}
Fermilab uses slip-stacking in the Recycler (and previously Main Injector) to double the proton bunch intensity it can deliver to experiments. The US Particle Physics community has come to a consensus that the Fermilab should upgrade its proton beam intensity in a cost-effective manner~\cite{P5}. To this end, the Fermilab Proton Improvement Plan-II~\cite{PIP} calls for an improvement in beam power from 700 kW (with slip-stacking) to 1.2 MW with an eye towards multi-MW improvements. The increase in proton intensity requires a commensurate decrease in the slip-stacking loss-rate to limit activation in the tunnel. A substantial improvement to either the Booster beam quality or stable slip-stacking bucket area would accomplish this objective. This Note describes the implications of both approaches.

\begin{figure}[htp]
	\centering
    \includegraphics[scale=0.35]{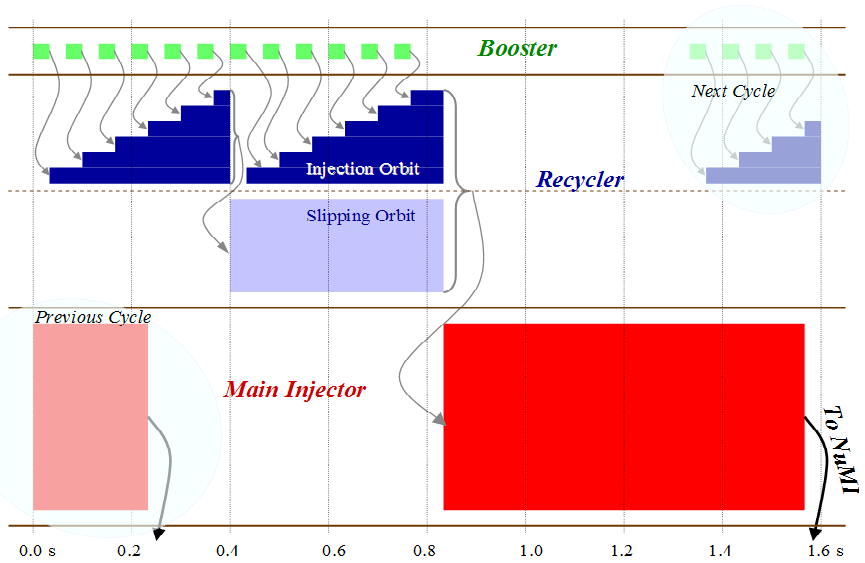}
  \caption{The green squares each represent an 82-bunch Booster batch injected directly into the Recycler. The first six batches in the Recycler slip past the next six (blue). Finally the twelve batches are extracted into the Main Injector and accelerated as one (red).}
  \label{SS}
\end{figure}

Slip-stacking allows two beams to accumulate in the same cyclic accelerator by using two RF cavities at near but distinct frequencies. Slip-stacking has been used at Fermilab since 2004 to nearly double the protons per ramp cycle~\cite{MacLachlan}\cite{SeiyaBC}. Slip-stacking in the Main Injector originally suffered significant beam-loading effects that were addressed through RF feedback and feedforward \cite{SeiyaB}. Previously slip-stacking took place in the Main Injector, but now takes place in the Recycler to avoid loading time. The complete ramp cycle with slip-stacking in the Recycler and a 15-Hz Booster cycle-rate is shown in Fig.~\ref{SS}.

The slipping rate of the buckets must be properly synchronized to the injection rate of new batches. The difference between the two RF frequencies must be equal to the product of the harmonic number of the Booster RF and the cycle rate of the Fermilab Booster. So for a Booster with a 15-Hz cycle-rate we have $$\Delta f = h_{B} f_{B} = 84 \times 15 \text{ Hz} = 1260 \text{ Hz}$$ and for a possible 20-Hz cycle-rate $$\Delta f = h_{B} f_{B} = 84 \times 20 \text{ Hz} = 1680 \text{ Hz}.$$
The difference in the frequency of the two RF cavities is related to the difference in momentum of the two beams by:
\begin{equation} \label{dd} 
\Delta \delta = \frac{\Delta f}{f_{rev} h \eta}
\end{equation}
where $h$ is the harmonic number of the Recycler and $\eta$ is the phase-slip factor of the Recycler (see Table~\ref{Param}). Consequently, the momentum difference between the two beams is $0.28\%$ for the 15-Hz Booster and $0.37\%$ for the 20-Hz Booster.

\begin{table}
\centering
\begin{tabular}{| l | l |}
\hline
Recycler Kinetic Energy ($E$) & 8 GeV \\
Recycler Reference RF freq. ($f$) & 52.8 MHz \\
Recycler Harmonic number ($h$) & 588 \\
Recycler Phase-slip factor ($\eta$) & -8.6*$10^{-3}$ \\
Maximum Recycler RF Voltage ($V$) & 2 $\times$ 150 kV \\
Booster harmonic number ($h_{B}$) & 84 \\ 
Booster cycle rate ($f_{B}$) & 15/20 Hz \\
Difference in Recycler RF freq. ($\Delta f$) & 1260/1680 Hz \\
\hline
Natural Booster Aspect Ratio & 3.00 MeV/ns \\
Natural Recycler Aspect Ratio (100 kV) & 1.06 Mev/ns \\
Natural Recycler Aspect Ratio (57 kV) & 0.80 MeV/ns \\
\hline
\end{tabular}
\caption{Recycler and Booster parameters used in analysis.}
\label{Param}
\end{table}

A 20-Hz Booster also reduces the time required to accumulate 12 batches in the Recycler, making more beam available for 8-GeV experiments~\cite{BooNE}\cite{mu2e}\cite{g-2}. For example, a 1.333 s MI cycle time would consume 9 Hz of the Booster's cycles, the additional available 8-GeV beam would increase from 6 Hz to 11 Hz.  Furthermore,  if the Main Injector ramp cycle is shortened to extract protons for LBNE at 60 GeV~\cite{LBNE}, then a 20-Hz Booster could deliver more beam to LBNE than a 15-Hz Booster.

\subsection*{Slip-stacking parameter}

Storing protons while slip-stacking is complicated by the fact that the two RF systems will interfere and reduce the stable bucket area. The further the buckets are away from each other in phase-space, the less interference there is. To quantify this, the  literature \cite{MacLachlan}\cite{Mills}\cite{Boussard} has identified the importance of the slip-stacking parameter
\begin{equation} \label{as} 
\alpha_{s} =\frac{\Delta f}{f_{s}}
\end{equation}
as the criterion for effective slip-stacking. $\Delta f$ is the frequency difference between the two RF cavities and $f_{s}$ is the single-RF synchrotron frequency $\displaystyle f_{s} = f_{rev} \sqrt{\frac{V h|\eta|}{2\pi \beta^{2} E}}$.

\begin{figure}[htp]
	\centering
    \includegraphics[scale=0.6]{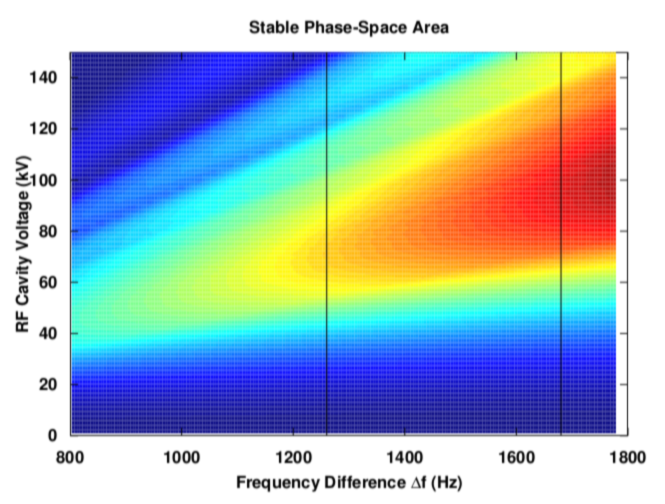}
  \caption{Stable phase-space area as a function of $\Delta f$ and $V$. Red indicates greater area, while blue is less.  The black-lines show the RF voltage tuning range for a 15-Hz Booster cycle-rate (left line) and a 20-Hz Booster cycle-rate (right line). The phase-space area with $\alpha_{s} > 8$ is extrapolated on this chart, but it is well constrained and far from the local maximums.}
  \label{VdF}
\end{figure}

The greater the slip-stacking parameter $\alpha_{s}$, the less the buckets interfere. Increasing the synchrotron frequency $f_{s}$ by increasing the voltage will increase the bucket height, but also increase the interference. So, for a fixed frequency difference $\Delta f$, there is a tradeoff encountered when tuning the voltage for maximum phase-space area. Fig.~\ref{VdF} shows how the $V$ and $\Delta f$ impact the stable bucket area.

\section*{Comparing Cycle-Rates and Injection Scenarios}

\begin{figure}[htp]
	\centering
    \includegraphics[scale=0.6]{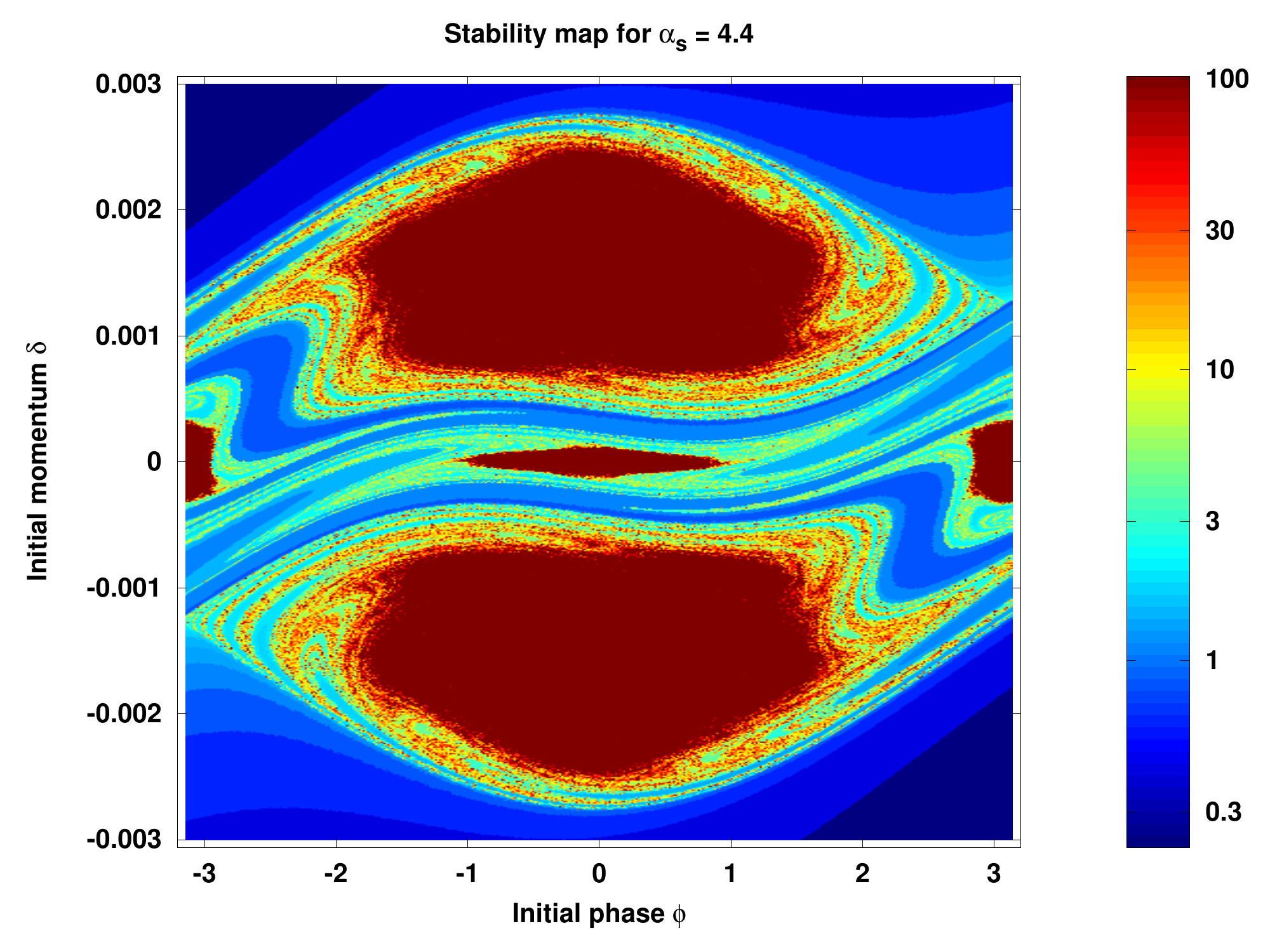}
  \caption{Stability of initial coordinates for $\alpha_{s} = 4.4$. The color corresponds to the number of synchrotron periods a particle with the corresponding initial coordinates survives before it is lost. The two large stable regions correspond to the higher and lower RF buckets where beam is injected and maintained.}
  \label{M}
\end{figure}

We map the stability of initial particle positions by integrating the equations of motion for each initial position. Each position is mapped independently and only the single particle dynamics are considered. The integration is iterated for 100 synchrotron periods. The stability of the particle is tested after every phase-slipping period. A particle is considered lost if its phase with respect to each of the first RF cavity, the second RF cavity, and the average of the two RF cavities, is larger than a certain cut-off (we used $3\pi/2$). Fig.~\ref{M} shows an example of such a stability map; the large stable regions at the top-center and bottom-center are the buckets used for slip-stacking and the interference effect is clearly evident.

\begin{figure}[htp]
\centering
 \begin{subfigure}
  \centering
  \includegraphics[scale=0.35]{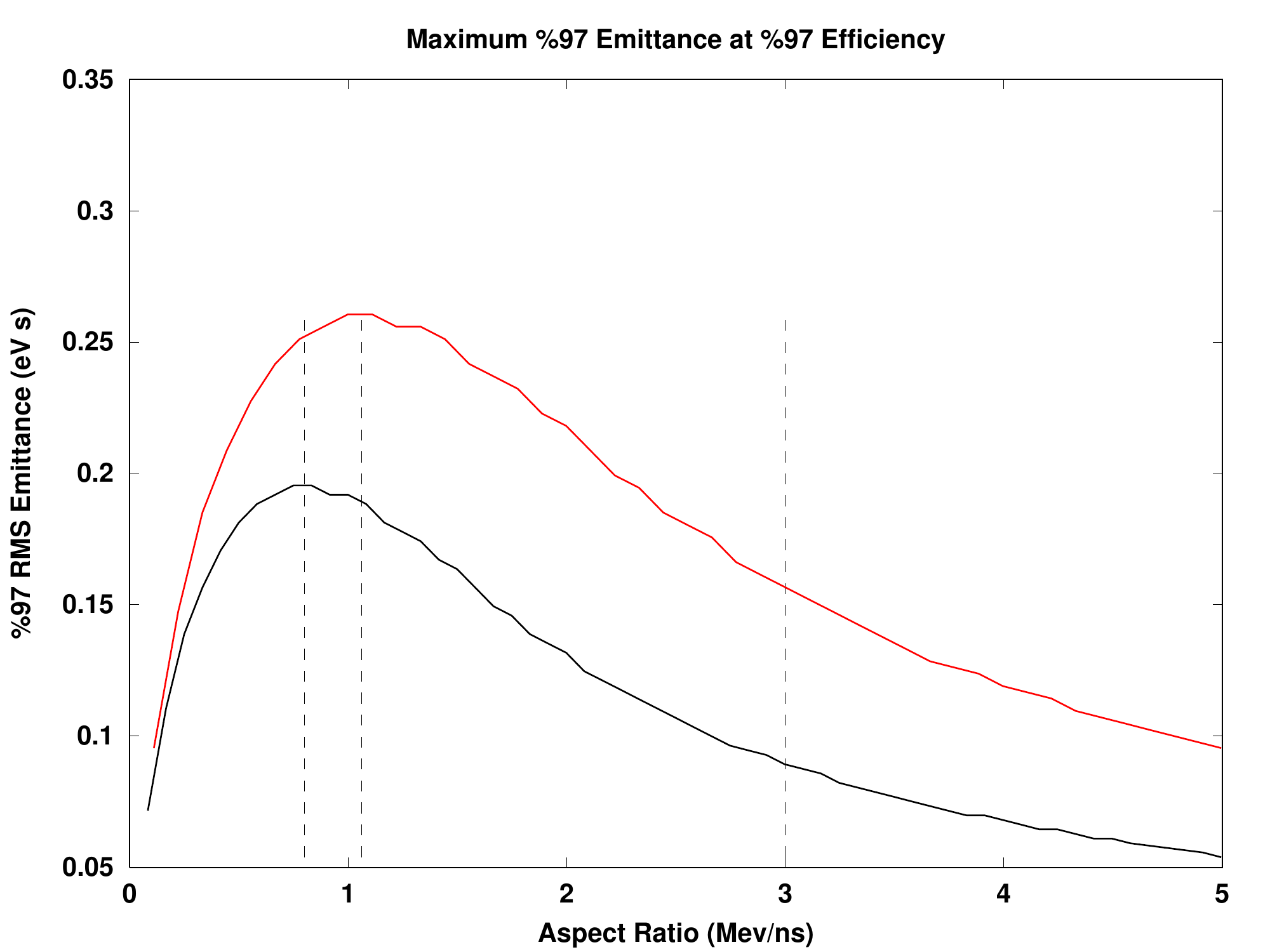}
 \end{subfigure}
 \begin{subfigure}
  \centering
  \includegraphics[scale=0.35]{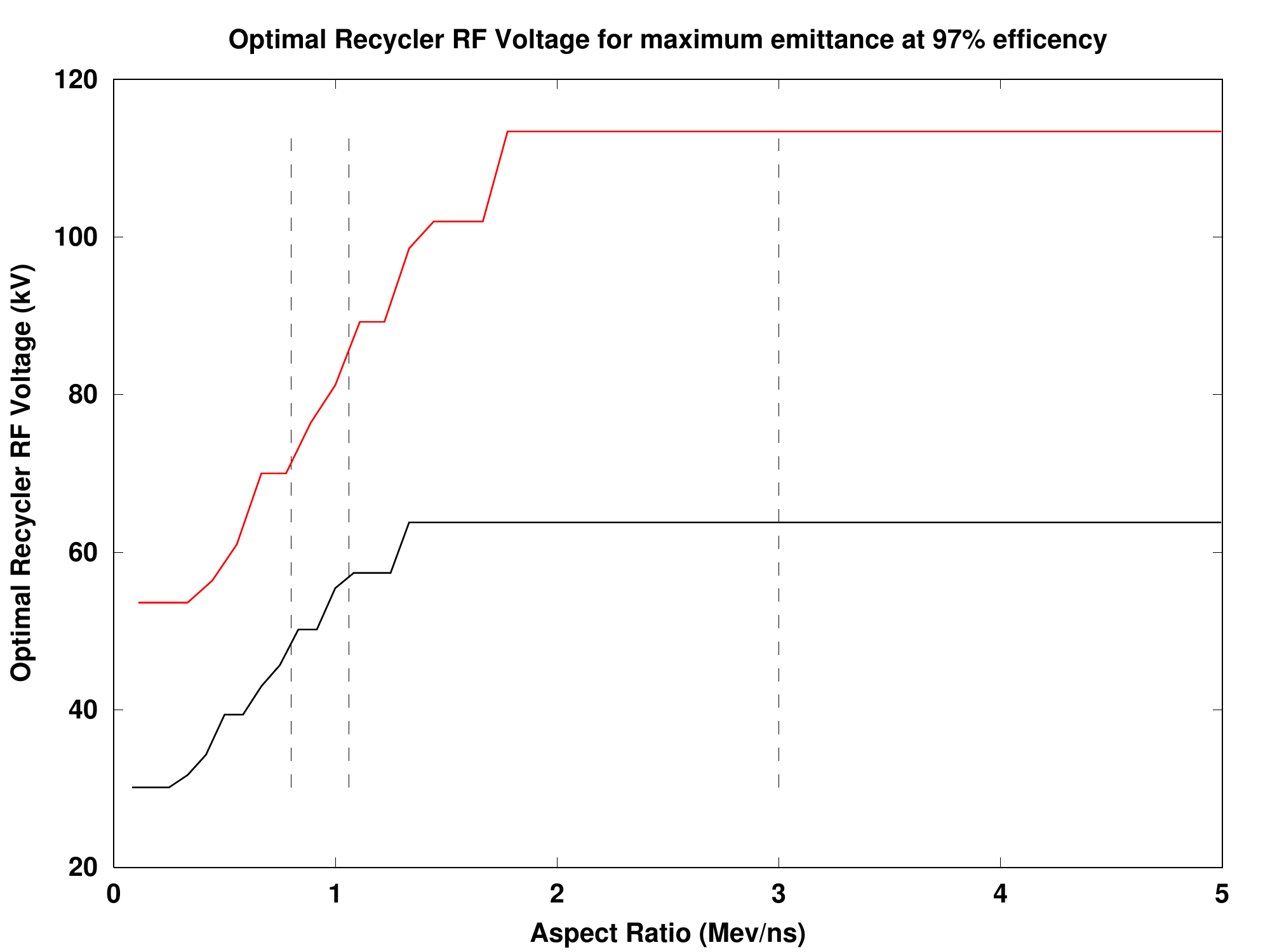}
 \end{subfigure}
 \caption{The bottom line (black) is for the 15-Hz Booster cycle-rate (status quo) and top line (red) is for 20-Hz Booster cycle-rate (hypothetical upgrade). The vertical dashed lines represent the natural aspect ratios given in Table~\ref{Param}. (a): Maximum emittance with 97\% efficiency (at an optimal value of $\alpha_{s}$) as a function of aspect ratio. 97\% Emittance, area of 97\% of the Gaussian distribution, is shown. (b): The optimal RF cavity voltage (for each cavity) as a function of aspect-ratio.}
 \label{AR}
\end{figure}

These stability maps can be used to analyze injection scenarios, by weighting the (appropriately scaled) stability maps according to a distribution that represents the number of incoming particles injected into that region of phase-space. We used this technique to identify the greatest longitudinal emittance an incoming Gaussian-distributed beam could have and still achieve 97\% injection efficiency at its optimal RF cavity voltage.  The longitudinal beam emittance is given in Eq.~\ref{lemit} below:
\begin{equation} \label{lemit}
\epsilon = \pi \sigma_{p}\sigma_{T},~\epsilon_{97\%} = 2.17^{2} \pi \sigma_{p}\sigma_{T}
\end{equation}
Fig.~\ref{AR}(a) shows this emittance as a function of aspect ratio and demonstrates the consequences of a mismatched injection into a slip-stacking bucket. The optimal RF cavity voltage as a function of aspect ratio  is shown in Fig.~\ref{AR}(b). These results were obtaining using parameter values shown in Table.~\ref{Param}.

The Fermilab Booster is matched to the Main Injector, not the Recycler, where the voltage is higher by a factor of 8 and therefore the natural aspect ratio is higher (narrower) by a factor of $\sqrt{8} \approx 2.83$. RF techniques such as paraphasing, voltage modulation~\cite{Yang}, and bunch rotation may be able to reduce the aspect ratio of the beam further. The current proton upgrade proposal, Proton Improvement Plan II (PIP-II)~\cite{PIP}, makes it clear that a 97\% slip-stacking efficiency is required to maintain current loss levels while increasing intensity.

\begin{figure}[htp]
	\centering
    \includegraphics[scale=0.6]{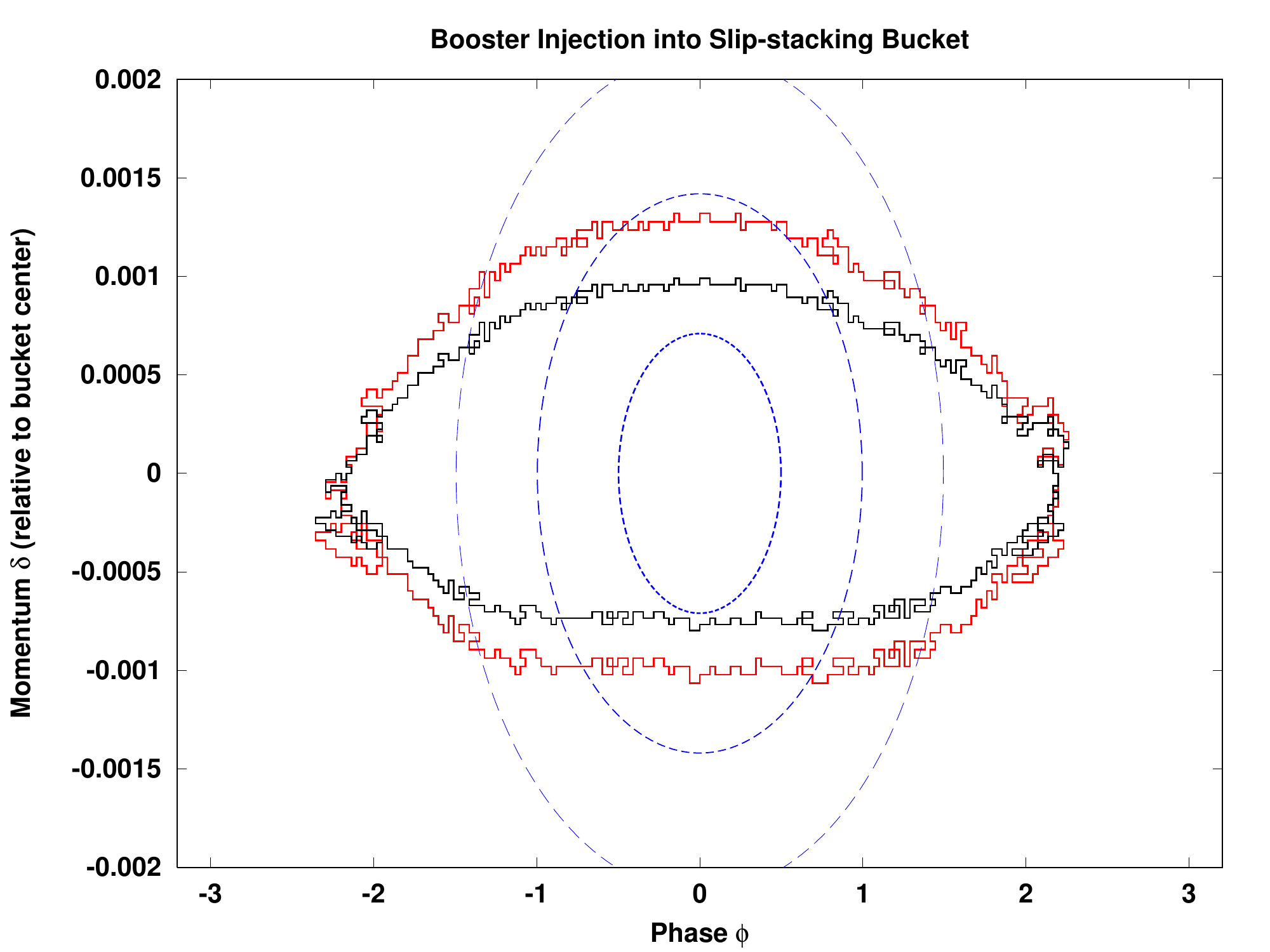}
  \caption{The shape of the slip-stacking Bucket is shown in black for the case of the 15-Hz Booster and in red for the case of the 20-Hz Booster. Both slip-stacking buckets are calculated for $\alpha_{s} = 5.5$ and optimized for bucket height. The three dashed blue lines represent $1\sigma$, $2\sigma$, and $3\sigma$ of a Gaussian distribution representing a typical Booster injection. In this case, the beam emittance is 0.1 eV$\cdot$s and the aspect ratio is 3 MeV/ns.}
  \label{Buckets}
\end{figure}

We examine the 97\% efficiency benchmark for both a 15-Hz Booster cycle-rate and a 20-Hz Booster cycle-rate. A 20-Hz Booster cycle-rate increases the RF frequency separation by a factor of 4/3 and therefore permits a 4/3 higher bucket height (for the same level of bucket interference). Consequently a 20-Hz Booster cycle-rate permits operation with either a significantly greater Booster emittance or injection efficiency. Fig.~\ref{Buckets} superimposes the Booster beam injection (natural aspect ratio without bunch rotation) for a 15-Hz Booster slip-stacking bucket and 20-Hz Booster slip-stacking bucket. Table~\ref{Emit} shows the improvement from a 20-Hz Booster cycle-rate expressed as a relaxation of Booster emittance limits. Table~\ref{Eff} shows the improvement from a 20-Hz Booster cycle-rate as greater efficiency.

\begin{table}
\centering
\begin{tabular}{| l | c | c |}
\hline
~ & 15 Hz & 20 Hz\\
\hline
Emittance Limit for 97\% Efficiency \& 3.00 MeV/ns & 0.089 eV$\cdot$s & 0.157 eV$\cdot$s \\ 
Emittance Limit for 97\% Efficiency \& 2.00 MeV/ns & 0.132 eV$\cdot$s & 0.218 eV$\cdot$s \\ 
\hline
\end{tabular}
\caption{Holding the aspect ratio and 97\% efficiency constant, the limits on Booster emittance are increased in a 20-Hz Booster.}
\label{Emit}
\end{table} 

\begin{table}
\centering
\begin{tabular}{| l | c | c |}
\hline
~ & 15 Hz & 20 Hz \\
\hline
Losses with 3.00 MeV/ns \& 0.08 eV$\cdot$s & {\bf 2.22 \%} & {\bf 0.30 \%} \\ 
Losses with 3.00 MeV/ns \& 0.1 eV$\cdot$s & 3.97 \% & {\bf 0.73 \%} \\ 
Losses with 3.00 MeV/ns \& 0.12 eV$\cdot$s & 5.95 \% & {\bf 1.38 \%} \\ 
Losses with 3.00 MeV/ns \& 0.18 eV$\cdot$s & 12.11 \% & 4.29 \% \\ 
\hline
Losses with 2.00 MeV/ns \& 0.08 eV$\cdot$s & {\bf 0.61 \%} & {\bf  0.04 \%} \\ 
Losses with 2.00 MeV/ns \& 0.1 eV$\cdot$s & {\bf 1.36 \%} & {\bf 0.14 \%} \\ 
Losses with 2.00 MeV/ns \& 0.12 eV$\cdot$s & {\bf 2.39 \%} & {\bf 0.33 \%} \\ 
Losses with 2.00 MeV/ns \& 0.18 eV$\cdot$s & 6.58 \% & {\bf 1.67 \%}  \\ 
\hline
\end{tabular}
\caption{Holding aspect ratio and emittance constant, the slip-stacking losses are dramatically reduced in a 20-Hz Booster. Bolded values pass the 97\% efficiency benchmark.  }
\label{Eff}
\end{table}

\begin{figure}[htp]
\centering
 \begin{subfigure}
  \centering
  \includegraphics[scale=0.35]{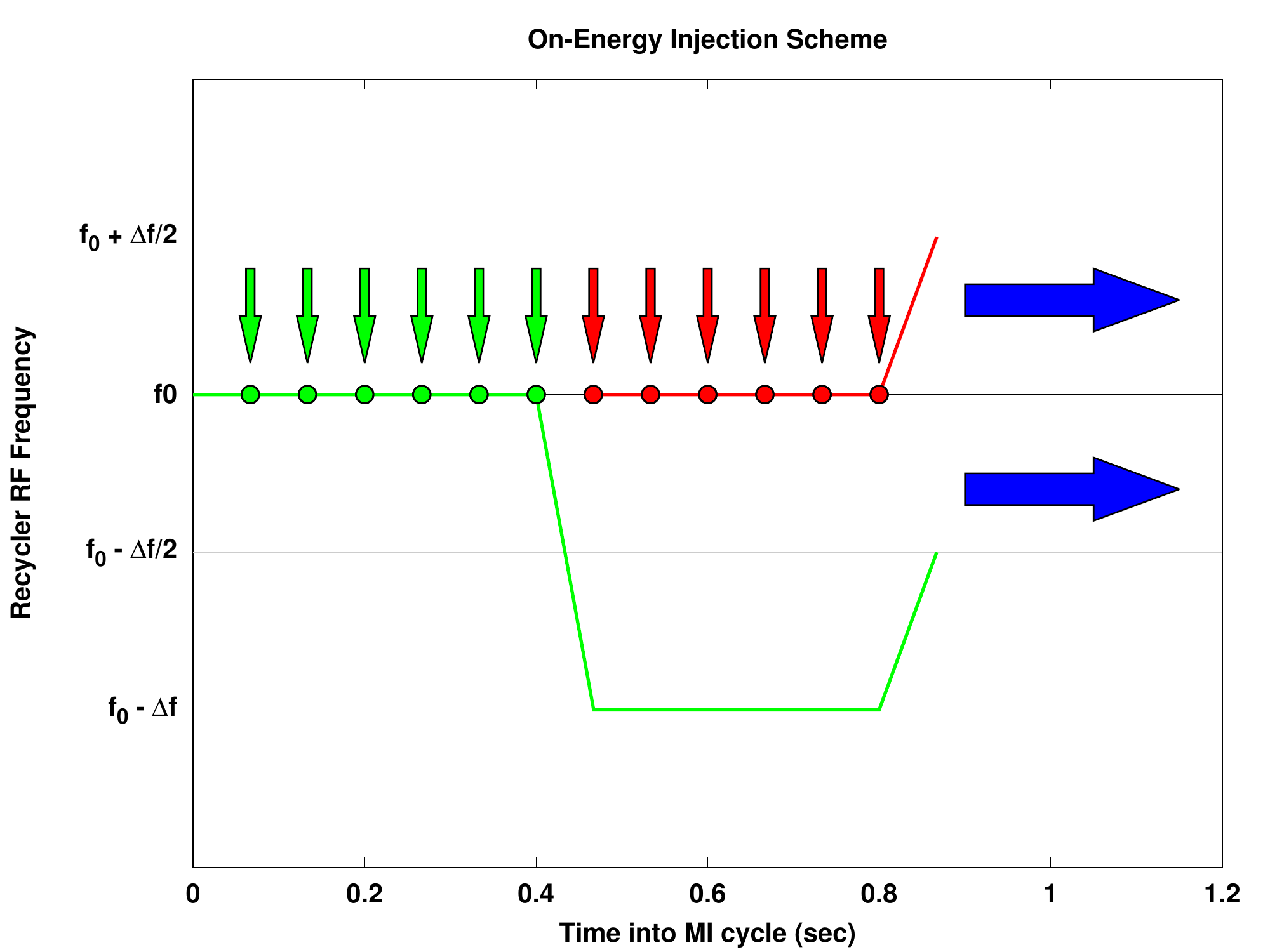}
 \end{subfigure}
 \begin{subfigure}
  \centering
  \includegraphics[scale=0.35]{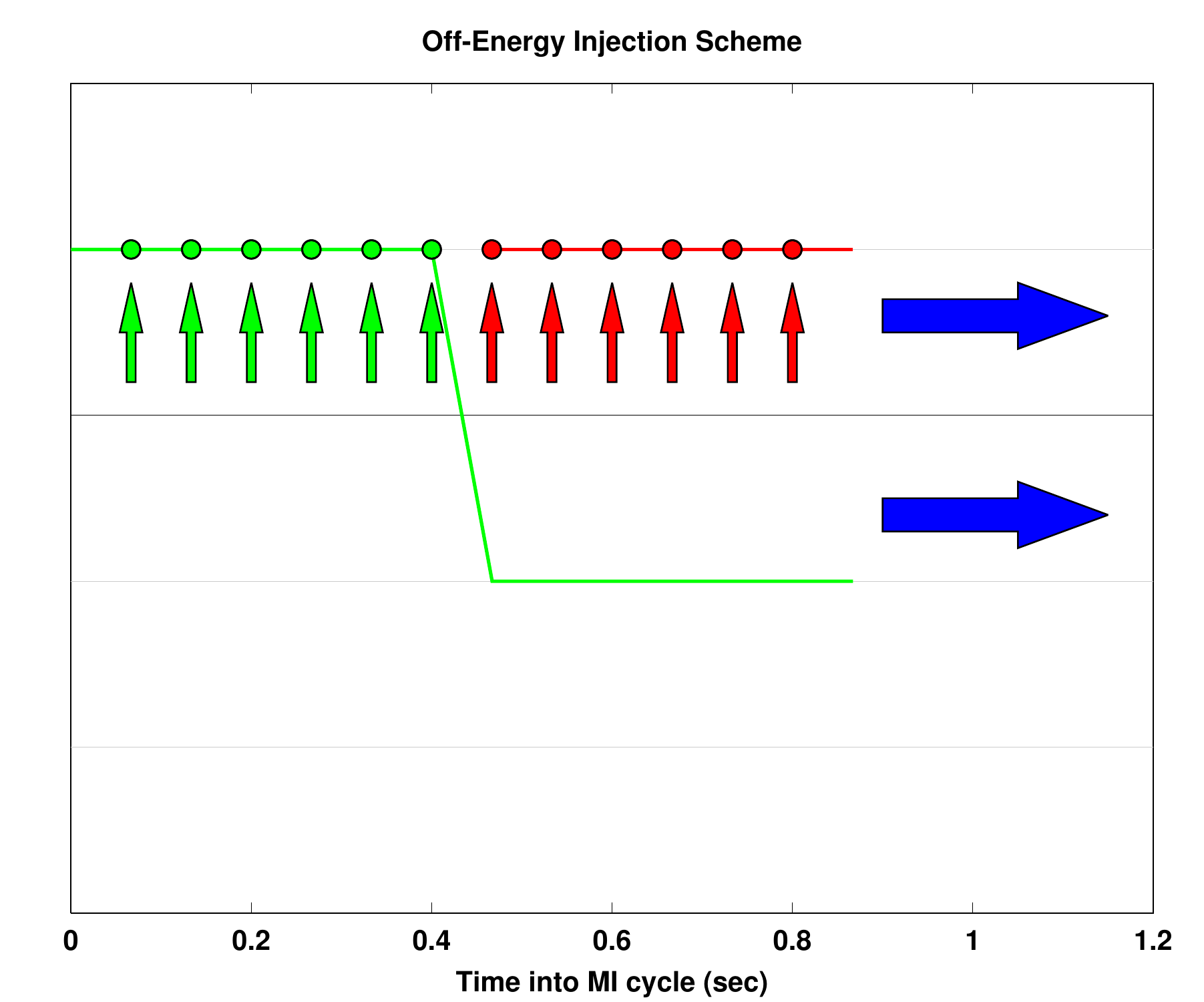}
 \end{subfigure}
 \caption{(a): The On-Energy injection scheme spans the frequencies $f_{0}+\Delta f/2$ to $f_{0}-\Delta f$. (b): Off-Energy injection scheme spans the frequencies $f_{0}+\Delta f/2$ to $f_{0}-\Delta f/2$.}
 \label{Inj}
\end{figure}

A 20-Hz Booster would best be implemented in conjunction with a change in the slip-stacking injection scheme to avoid encountering limits in momentum aperture. See \cite{Kourbanis} and \cite{Scott} for recent measurements of the Recyler momentum aperture. It should be noted that this momentum aperture is limited by the dynamic aperture, which means that it is sensitive to chromaticity and betatron tuning; it is approximately half of the phsyical aperture which may be achieved with improvements to the lattice. The total momentum range used during slip-stacking is shown in Table~\ref{Mom}. Because the 20-Hz Booster requires greater RF frequency separation, the total momentum used in any injection scheme would increase. But as Table~\ref{Mom} indicates, switching from the ``On-Energy'' injection with a 15-Hz Booster (status quo) to ``Off-Energy'' injection 20-Hz Booster (proposed) is actually a net decrease in the total momentum usage. These two injection schemes are depicted in Fig.~\ref{Inj}.  


In the ``On-Energy'' injection scheme (see Fig.~\ref{Inj}(a)), the extraction energy from the Booster is the injection energy into the Recycler.  The frequencies of the Recycler RF cavities move to ensure the injection and extraction is simple, but at the cost of greater momentum usage. In the ``Off-Energy'' injection scheme (see Fig.~\ref{Inj}(b)), the Recycler must be tuned to extract at a momentum $\Delta \delta /2$ lower or higher than the momentum of the beam injected into the Main Injector (\cite{NOvA}, p.\ 8-109). The advantage offered by this alternate injection scheme is that only $\Delta \delta$ and the full bucket height must be accommodated, rather than the $(3/2) \Delta \delta$ and the full bucket height required by the On-Energy injection scheme. Eq.~\ref{dd} relates the frequency difference with the momentum difference.

\begin{table}
\centering
\begin{tabular}{| l | c | c |}
\hline
~ & 15 Hz & 20 Hz \\
\hline
Momentum Usage with $\pm$ 12 MeV \& ``On-Energy'' Injection & 0.72 \% & 0.86 \% \\ 
Momentum Usage with $\pm$ 12 MeV \& ``Off-Energy'' Injection & 0.58 \% & 0.67 \% \\ 
\hline
Momentum Usage with $\pm$ 8 MeV \& ``On-Energy'' Injection & 0.63 \% & 0.76 \% \\ 
Momentum Usage with $\pm$ 8 MeV \& ``Off-Energy'' Injection & 0.48 \% & 0.57 \%  \\ 
\hline
Momentum Usage with $\pm$ 4 MeV \& ``On-Energy'' Injection & 0.52 \% & 0.66 \% \\ 
Momentum Usage with $\pm$ 4 MeV \& ``Off-Energy'' Injection & 0.38 \% & 0.47 \% \\ 
\hline
\end{tabular}
\caption{Holding momentum range and the injection scheme constant, the 20-Hz Booster cycle-rate requires greater momentum aperture in the Recycler. However, the greater usage can be entirely compensated by using the ``Off-Energy'' injection scheme. The momentum range represents the complete momentum range of the Booster beam. Fig.~\ref{Inj} depicts the two injection schemes.}
\label{Mom}
\end{table}

The gains in slip-stacking efficiency under the 20-Hz Booster scenario also require an increase in RF cavity voltage (see Fig.~\ref{AR}(b)). The ideal RF cavity voltage increases from 64 kV to 114 kV, which is a factor of $(4/3)^{2} \approx 1.78$. The duty factor may also decrease (by no more than $3/4$) in the case of a 20-Hz Booster; the power dissipation would increase by at least $(4/3)^3 \approx 2.37$. The maximum Recycler RF voltage is 150 kV and the maximum Recycle RF power is 150 kW, according to \cite{Madrak}. The possibility of the Recycler RF cavities overheating would have to be investigated.

\end{document}